# Study of Crystal-field Effects in Rare-earth (RE) – Transition-metal Intermetallic Compounds and in RE-based Laser Crystals

NICOLA MAGNANI

*XV Ciclo – Dottorato di Ricerca in Fisica*
*Dipartimento di Fisica, Università di Parma*
*Parco Area delle Scienze, 7/A – 43100 Parma, Italy*
*magnani@fis.unipr.it*

## 1. INTRODUCTION

Rare-earth (RE) based compounds and alloys are of great interest both for their fundamental physical properties and for applications (*e.g.* for producing new-generation solid-state laser and high-performance magnetic materials). In order to tailor the required compounds for a specific task, one must be able to predict the energy level structure and transition intensities for any magnetic ion in any crystalline environment.

The crystal-field (CF) analysis is one of the most powerful theoretical methods to deal with the physics of magnetic ions. The problem of analyzing the microscopic origin of the CF splitting, which is observed in the optical spectra of RE-based crystals, without relying only on the simple electrostatic point-charge model has been studied from the 1950's. One of the most known and brilliant solutions is Newman's Superposition Model (Newman & Ng, 2000); this semiphenomenological analysis cannot resolve the exact electronic structure of the magnetic centers, but it provides definite information on the correlation of the CF potential with the geometry around the active site. Today, this technique is widely used for the data analysis of optical spectroscopy experiments, in particular to analyze the pressure dependence of the spectral lines or to study local lattice distortions; whereas the latter information is deduced from X-ray structural determination in pure crystals, in low-concentration doped materials the crystallographic approach is not suitable and the spectral analysis becomes a key tool. It must be mentioned that, although these theoretical methods have become quite standard, they are still the object of considerable debate; in particular the possibility to use the data obtained with the Superposition Model in a given system to predict the energy spectra of other similar compounds, which can be very useful for optical tayloring, is frequently discussed in recent literature papers.

Apart from optical spectroscopy, CF analysis is also useful in many other fields. For example, a significant breakthrough in the figures of merit of permanent magnets was achieved during the 1970's by studying various alloys of RE and transition-metal elements. In fact, the inclusion of RE ions in soft ferromagnets can result in a strong anisotropy and coercitivity, due to the fact that the non-spherical CF potential gives rise to a preferential direction for the RE magnetic moment. In addition to the large magnetization and high Curie temperature of transition metals such as cobalt, this feature makes materials such as $Sm_2Co_{17}$ or $Nd_2Fe_{14}B$ the best candidates as basic constituents of permanent magnets for industrial applications.

## 2. CRYSTAL-FIELD ANALYSIS OF RARE-EARTH DOPED $BaY_2F_8$

Yttrium fluorides doped with trivalent RE ions are good candidates as active materials for new-generation lasers and scintillators; since the RE substitution occurs at the unique $Y^{3+}$ site, charge compensation is not required and the dopant level can be widely varied. The knowledge of the spectroscopic properties of these compounds is crucial in order to define their perspectives of utilization. In the present work, high-resolution FTIR spectroscopy was applied to study the monoclinic Ce-, Dy-, and Er-doped BaY$_2$F$_8$ single crystals, which were grown by means of the Czochralsky method. Relatively low RE concentrations were chosen in order to avoid the additional lines due to the presence of RE clusters which may appear in the spectra of highly-doped samples (Baraldi et alii, 2001). The experimentally detected peaks in the absorption spectra were then assigned to the corresponding transitions in order to obtain the energy level scheme for the three considered RE ions.

In order to analyze the data, a theoretical single-ion Hamiltonian accounting for free-ion interactions and crystal-field potential was considered:

$$\hat{H} = E_{av} + \sum_k F^k \hat{f}_k + \zeta \hat{H}_{SO} + \alpha \hat{G}(R_3) + \beta \hat{G}(G_2) + \gamma \hat{G}(R_7)$$
$$+ \sum_i T^i \hat{t}_i + \sum_j M^j \hat{m}_j + \sum_k P^k \hat{p}_k + \sum_{k,q} B_k^q \hat{C}_q^{(k)}. \qquad (1)$$

The definitions of operators and coefficients in Eq. (1) are standard and extensively discussed in the literature (Carnall et alii, 1989; Wybourne, 1965). In order to use a reasonable number of degrees of freedom, no more than 5 of the 19 free-ion parameters were freely varied during the fitting procedure; literature data for similar compounds were used to complete the set. Conversely, all the 14 crystal-field parameters allowed by the C$_2$ symmetry of the Y site were varied. The experimental and calculated energy levels are in good agreement, with a deviation $\sigma$ between 6 and 8 cm$^{-1}$ for all the considered RE dopants.

The obtained CF parameters (Magnani et alii, 2002) were then analyzed by means of Newman's Superposition Model, which is based on the hypothesis that the CF potential at the RE site is the sum of axially symmetric individual contributions from each ligand, the symmetry axis being the straight line which links the considered ligand to the central ion. In this framework, the CF parameters can be expressed as

$$B_k^q = N_k^q \langle r^k \rangle \sum_\ell \overline{A}_k(R_0) \left( \frac{R_0}{R_\ell} \right) K_k^q(\theta_\ell, \varphi_\ell), \qquad (2)$$

where $K_k^q(\theta,\varphi)$ are the so-called *coordination factors* (Newman, 1971); it has been supposed that the intrinsic parameters $A_k(R)$ have a power-like dependence on the ligand distance.

The SPM parameters which were obtained by this analysis are similar to those given in the literature for RE$^{3+}$ ions in other F-based compounds such as LaF$_3$; the general agreement achieved for a series of four different RE dopants in the same host is a remarkable proof of the validity of the Superposition Model for this class of compounds. In particular, relevant information was gained regarding the modification of the local structure around substituted RE$^{3+}$ ions; the calculated levels do not match their experimental counterparts without taking into account an anisotropic distortion of the F$^-$ ligand cage for heavy rare earths, roughly shown in Fig. (1).

## 3. *J*-MIXING LINEAR THEORY OF MAGNETOCRYSTALLINE ANISOTROPY

In principle, the axial part of the magnetic anisotropy in a RE-TM (TM = transition-metal) can be written as

$$E_A = \sum_{n=0}^\infty K_n \sin^{2n} \theta, \qquad (3)$$

where $\theta$ is the angle between the chosen axis and the magnetization vector; the terms odd in $\theta$ do not appear in Eq. (3) since $E_A$ must not change when the magnetization direction is reversed. The microscopic mechanism which is responsible for the formation of the strong RE anisotropy in these compounds may be studied by the single-ion Hamiltonian

$$\hat{H}_{RE} = \Lambda \mathbf{L} \cdot \mathbf{S} + 2\mu_B \mathbf{H}_{ex} \cdot \mathbf{S} + \sum_{k,q} B_k^q \hat{C}_q^{(k)}. \qquad (4)$$

The linear theory of magnetocrystalline anisotropy consists in calculating the first-order contribution to the free energy due to the RE ion within the ground multiplet, by defining the generalized Brillouin functions (GBFs) as

$$\tilde{B}_J^{(k)}(x) = \frac{\langle \hat{C}_0^{(k)} \rangle}{J^k} = \frac{\sum_{M=-J}^{+J} \langle J,M | \hat{C}_0^{(k)} | J,M \rangle \exp(-xM/J)}{J^k \sum_{M=-J}^{+J} \exp(-xM/J)} \quad (5)$$

This approach leads to a simpler expression for $E_A$, where all the anisotropy terms of eight order and higher vanish, leading to

$$E_A = K_1 \sin^2\theta + K_2 \sin^4\theta + K_3 \sin^6\theta, \quad (6)$$

and the three nonzero anisotropy constants can be written in analytical form (Kuz'min, 1992). Unfortunately, this method cannot be used to analyze the data regarding compounds of light RE ions such as $Sm^{3+}$, since the mixing of states belonging to different multiplets (known as $J$ mixing) is neglected.

In this work, we propose the use of a perturbative technique (Slichter, 1990) in order to include the so far neglected $J$-mixing effects in the linear theory; a similar approach has recently been used to study the mixing between different spin states in molecular clusters such as $Cr_8$ (Carretta et alii, 2003). We have shown that it is possible to find an Hermitian operator $\Omega$ such that the transformed Hamiltonian

$$\hat{H}'_{RE} = e^{-i\Omega} \hat{H}_{RE} e^{i\Omega} \quad (7)$$

has very small matrix elements in the off-diagonal blocks. It can be demonstrated that the formalism of the isolated-multiplet description can be retained while correctly taking into account the $J$-mixing effects on the anisotropy, by simply adding a term to the Hamiltonian:

$$\hat{H}_{mixing} = \frac{2\mu_B H_{ex}}{\Delta_{SO}} \sum_{k,k',q} B_k^q \hat{M}_q^{(k')} \quad (8)$$

where $\Delta_{SO}$ is the spin-orbit splitting between the ground and the first excited multiplet, and $M_q^{(k')}$'s are linear combinations of $C_q^{(k)}$ tensor operators with appropriate coefficients. Once this is done, the linear theory can be applied in presence of $J$ mixing, by simply interpreting Eq. (8) as an effective crystal-field potential (Magnani et alii, 2003). It is found that Eq. (6) is still valid, although the temperature dependence of the anisotropy constants changes significantly. It must be noted that the main difference between $H_{mixing}$ and $H_{CF}$ is that the latter only contains even-order operators, while in the former operators of odd order are present; this is necessary to guarantee time-reversal invariance, as $H_{mixing}$ depends linearly on the exchange field. This leads to the presence of odd-order GBFs in the analytical expressions of the $K_n$; an example of this peculiar temperature dependence due to $J$ mixing is the basal-plane anisotropy constant of $Sm_2Co_{17}$ (Magnani et alii, 2000).

## 4. MAGNETIC FIELD-INDUCED TRANSITIONS IN $Pr_2(Co,Fe)_{17}$ COMPOUNDS

Praseodymium-based intermetallics are particularly interesting from the fundamental point of view, since peculiar features such as the conical arrangement of the easy magnetization direction in $Pr_2Co_{17}$ (Verhoef et alii, 1988) and the occurrence of two distinct field-induced magnetic transitions in $Pr_2Fe_{17}$ (Kou et alii, 1998) have been reported. For this reason, we performed a complete study of the anisotropy and field-induced transitions in $Pr_2Co_{17-x}Fe_x$ (Pareti et alii, 2002). 16 samples with different composition were prepared by arc melting, along with the corresponding $Y_2Co_{17-x}Fe_x$ samples which are used as "blank" in order to separate the effects of $3d$ and $4f$ electrons on the anisotropy. The singular point detection technique (Asti & Rinaldi, 1972) has been used to measure the anisotropy field and the critical field of first-order magnetization processes (FOMPs) in all compounds. It is worth to recall that a FOMP is a transition associated to the irreversible rotation of the magnetization vector $\mathbf{M}_S$ between two inequivalent states, which occurs at a critical field value $H_{cr}$. Type-1 and 2 transitions refer respectively to jumps of $\mathbf{M}_S$ to saturation (*i.e.* parallel to the field direction) or to an intermediate position. While Y compounds do not show any field-induced transitions, the diagram in Fig. (2) clearly shows that all Pr compositions with x ≠ 0 display at least a FOMP; some of these transitions are present up to room temperature, where the high-order anisotropy contributions required to have a FOMP are usually negligible.

Since the presence of a double transition for $x \geq 13.6$ is incompatible with the FOMP description which can be given by means of Eq. (6) (Asti & Bolzoni, 1980), in order to analyze the experimental data with a theoretical crystal-field model it was necessary to perform numerical calculations based on Eq. (4). Two main results were obtained:

(i) The Pr-sublattice contribution is essential for the development of the FOMP transitions. On the other hand, the $3d$-sublattice Co-Fe composition markedly affects the above contribution, which is large and negative (planar) at high Co content while it abruptly changes sign at higher Fe concentrations. A critical behaviour of the leading crystal-field parameter $B_2^0$ is found in the composition region $x \approx 5$; this also leads to significant effects in the magnetostriction (Magen et alii, 2003).

(ii) The double-FOMP behaviour in Fe-rich compounds cannot be explained if the RE ions occupy only one single site, as reported in the literature. By performing numerical simulations of magnetization curves with a phenomenological two-RE-sublattice model, we were able to reproduce qualitatively the experimentally observed behaviour. In fact, X-ray diffraction experiments performed on several $Pr_2Fe_{17}$ single crystals pointed out the presence of stacking faults in the crystal structure (obverse-reverse twinning domains). The modified Fe environment experienced by two independent Pr ions lying on a twinning plane gives rise to a second, magnetically inequivalent, RE sublattice.

## 5. CONCLUSIONS

The experimental values of the energy levels of several RE-doped BaYF crystals were fitted to a single-ion Hamiltonian accounting for free-ion and crystal-field interactions. Our aim was to obtain a microscopic description of the crystal-field potential in terms of the geometry of the ligands around the optically active site, which may be useful in view of possible applications of this material. A unified picture for the four considered dopants was obtained by assuming a distortion of the $F^-$ ligand cage; from the theoretical point of view, the possibility to account for all the crystal-field parameters in a low-symmetry system by means of the Superposition Model is a relevant contribution to the debate on the limits of its applicability.

A more detailed theoretical model is required to account for the magnetic properties of intermetallic alloys. By means of a perturbative technique, we obtained analytical formulae to study the behaviour of compounds with strong $J$ mixing; this is crucial to explain some features such as the presence of basal-plane anisotropy in $Sm_2Co_{17}$. Our approach to $J$-mixing problems has several advantages over the previously used numerical methods; in particular, it can provide a deeper insight into the physics which governs these systems.

In the case of the intermetallic compounds $Pr_2(Co,Fe)_{17}$, several types of magnetic transitions (such as field-induced magnetization processes) have been experimentally detected; most of them can be reproduced by the single-ion model, giving an insight of some peculiar microscopic properties of this system (*e.g.* the leading crystal-field parameter has a threshold dependence on the Co-Fe composition). The double field-induced transition detected for iron-rich compounds cannot be explained without assuming that two inequivalent RE sublattices are present; this was verified by X-ray structural determination, which also unveiled novel features of these compounds which were not present in the literature (Calestani et alii, 2003).

## ACKNOWLEDGEMENTS

I am very grateful to Prof. Giuseppe Amoretti, Supervisor of this Thesis, for his continuous support; to Dr. Luigi Pareti and all the Magnetic Materials group at the IMEM – CNR Institute, where all the magnetic measurements were performed; to Prof. Rosanna Capelletti and Dr. Andrea Baraldi (Dipartimento di Fisica, Università di Parma) for the optical measurements; to Prof. Gianluca Calestani (Dipartimento di Chimica G.I.A.F., Università di Parma) for X-ray diffraction measurements; to Prof. Ricardo Ibarra, Dr. Pedro Algarabel, Dr. Luis Morellon and Cesar Magen (ICMA – CSIC, Zaragoza) for the magnetostriction measurements.

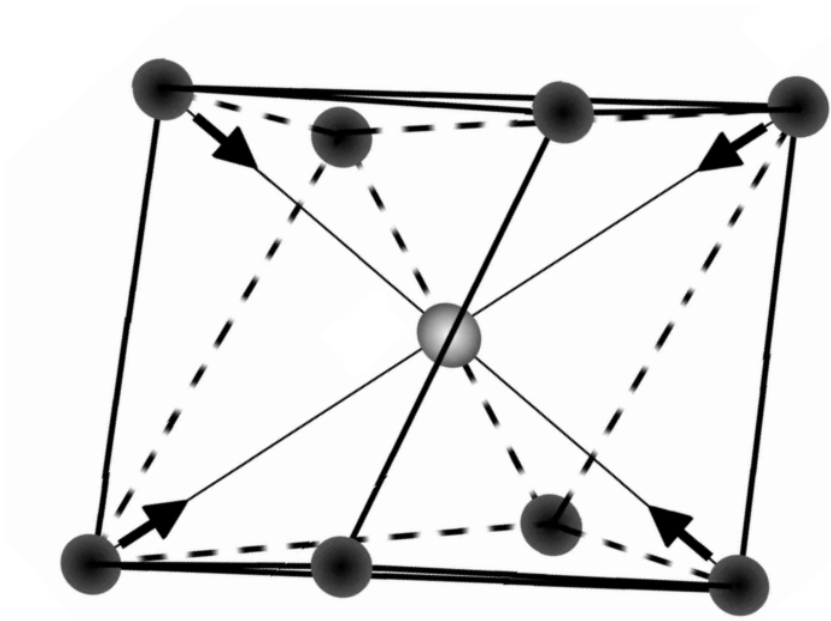

**Fig. 1:** The model anisotropic distortion of the $F_8$ polyhedron. The arrows indicate the four involved $F^-$ ions. The grey ball represent the $RE^{3+}$ ion, while dark balls are the eight surrounding $F^-$ ligands.

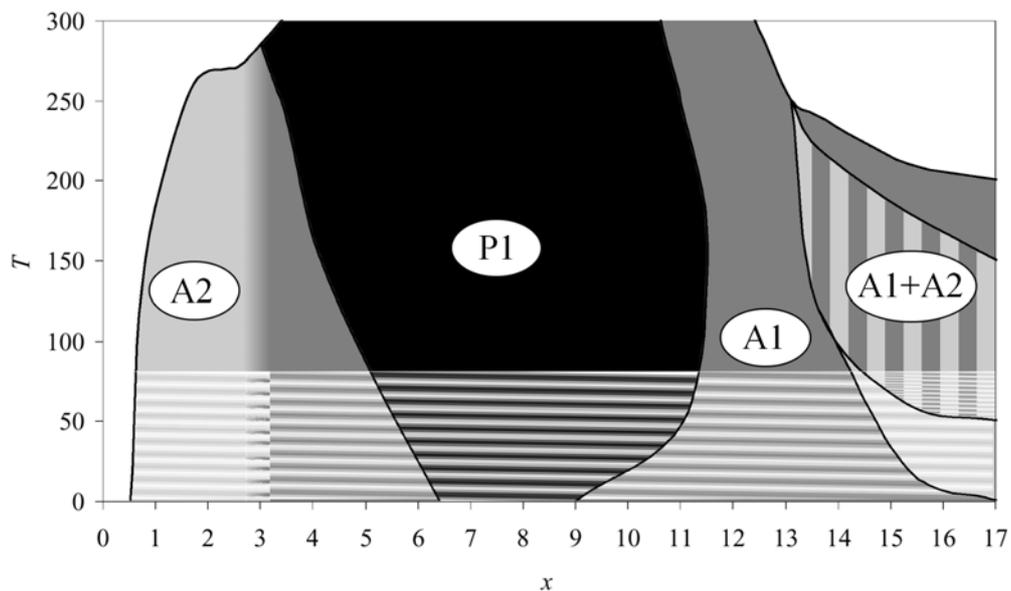

**Fig. 2:** FOMP diagram of $Pr_2Co_{17-x}Fe_x$. Black zone: P1-type FOMP. Dark grey zone: A1-type FOMP. Light grey zone: A2-type FOMP. White zones: no field-induced transitions detected. Stripes of different colours indicate the simultaneous presence of two different transitions. Temperatures below 78 K were not reached experimentally, and the corresponding shaded area is an extrapolation based on literature data.